\documentstyle[aps,twocolumn,epsf]{revtex}
\begin{document}
\draft
\title{Dynamic black-hole entropy}
\author{Sean A. Hayward,\thanks{\tt hayward@yukawa.kyoto-u.ac.jp}
Shinji Mukohyama\thanks{\tt mukoyama@yukawa.kyoto-u.ac.jp}}
\address{Yukawa Institute for Theoretical Physics, Kyoto University,
Kitashirakawa, Sakyo-ku, Kyoto 606-8502, Japan}
\author{M. C. Ashworth\thanks{\tt ashworth@th.phys.titech.ac.jp}}
\address{Department of Physics, Tokyo Institute of Technology,
Oh-okayama, Meguro, Tokyo 152-8551, Japan}
\date{2nd October 1998}
\maketitle

\begin{abstract}
We consider two non-statistical definitions of entropy
for dynamic (non-stationary) black holes in spherical symmetry.
The first is analogous to the original Clausius definition
of thermodynamic entropy:
there is a first law containing an energy-supply term
which equals surface gravity times a total differential.
The second is Wald's Noether-charge method,
adapted to dynamic black holes by using the Kodama flow.
Both definitions give the same answer for Einstein gravity:
one-quarter the area of the trapping horizon.

\end{abstract}

It is widely believed that black holes possess a gravitational entropy,
given in Einstein gravity by $A/4$, where $A$ is the black-hole area
and units are such that $c=G=\hbar=k=1$.
The original reasoning stemmed from two discoveries:
(i) Hawking's result\cite{H} that for stationary black holes,
quantum fields radiate with a thermal spectrum at temperature $\kappa/2\pi$,
where $\kappa$ is the surface gravity of the black hole;
and (ii) previously discovered properties of stationary black holes\cite{C}
which are analogous to the laws of thermodynamics,
in particular a first law of the form $\delta m=\kappa\delta A/8\pi$
plus work terms, where $\delta$ is a perturbation
of a stationary black-hole solution of mass $m$.
This has been called the first law of black-hole statics
to stress that it concerns stationary black holes\cite{1st}.
Most discussions of black-hole entropy concern the stationary case,
which is analogous to thermostatics rather than thermodynamics.
Since entropy is a fundamental quantity in thermodynamics
and not just thermostatics, one expects that
dynamic (i.e.\ non-stationary) black holes should also possess an entropy.

This leads to two questions about potential generalizations of the above facts.
(i) Do dynamic black holes have a local Hawking temperature in some sense?
(ii) Is there a first law of black-hole dynamics?
The latter question was recently answered in Ref.\cite{1st}
in spherical symmetry.
The method is based on earlier work
which defined dynamic black holes in terms of trapping horizons
and derived a corresponding second law of black-hole dynamics\cite{bhd}.
Moreover,
this first law of black-hole dynamics is encoded in a unified first law
which also encodes a first law of relativistic thermodynamics.
A possible answer to the former question was also suggested,
by a local definition of surface gravity for a dynamic black hole.
Whether this really determines a local Hawking temperature is still unknown.

It is generally thought that
black-hole entropy should have a statistical origin,
presumably in a quantum theory of gravity.
This is, of course, due to the definition of entropy in statistical mechanics.
However, it should be remembered that
the original concept of entropy was not statistical\cite{T}.
The original argument of Clausius was that, in a cyclic reversible process,
the total heat supply $\delta Q$ divided by temperature $\vartheta$
should vanish.
Thus in any reversible process, $\delta Q/\vartheta$
should be the total differential $dS$ of a state function $S$, the entropy.
Moreover, in irreversible processes,
there should be a second law $dS\ge\delta Q/\vartheta$.
The heat supply also occurs in a first law $dU=\delta Q+\delta W$,
where $U$ is the internal energy and $\delta W$ the work being done.
These are basic laws of thermodynamics as stated in typical textbooks and
originally formulated by Clausius before the invention of statistical mechanics.
In this article, we argue that there is a similar concept
of entropy for dynamic black holes,
suggested by the mathematical structure of the unified first law:
it contains an energy-supply term
which equals surface gravity times a total differential.

The relevant quantities and equations in spherical symmetry
may be summarized as follows.
The area $A$ or areal radius $r=\sqrt{A/4\pi}$ of the spheres of symmetry
determines the 1-form
\begin{equation}
k={*}dr
\end{equation}
where $d$ is the exterior derivative and $*$ is the Hodge operator
of the two-dimensional space normal to the spheres of symmetry.
Henceforth, 1-forms and their vector duals
with respect to the space-time metric will not be distinguished.
Then $k$ is the divergence-free vector introduced by Kodama\cite{K},
which generates a preferred flow of time
and is a dynamic analogue of a stationary Killing vector\cite{1st}.
The active gravitational energy or mass is
\begin{equation}
E=(1-dr\cdot dr)r/2
\end{equation}
where the dot denotes contraction.
Misner \& Sharp\cite{MS} originally defined $E$
and Ref.\cite{ge} described its physical properties.
The dynamic surface gravity
\begin{equation}
\kappa={*}dk/2
\end{equation}
was defined in Ref.\cite{1st}
by analogy with the standard definition of stationary surface gravity.
This reference also introduced two invariants of the energy tensor $T$:
the energy density (work density)
\begin{equation}
w=-\hbox{tr}\,T/2
\end{equation}
and the energy flux (localized Bondi flux)
\begin{equation}
\psi=T\cdot dr+wdr
\end{equation}
where tr denotes the the two-dimensional normal trace.
One may say that $(A,k)$ are the basic kinematic quantities,
$(E,\kappa)$ the gravitational quantities
and $(w,\psi)$ the relevant matter quantities.
Instead of $\psi$ one may also use
the divergence-free energy-momentum vector\cite{1st,K,ge}
\begin{equation}
j={*}\psi+wk.
\end{equation}
Finally, the relevant components of the Einstein equation are\cite{1st}
\begin{equation}
E=r^2\kappa+4\pi r^3w
\end{equation}
and\cite{ge}
\begin{equation}
Aj={*}dE.
\end{equation}
The latter may be rewritten as
\begin{equation}
dE=A\psi+wdV
\end{equation}
where $V={4\over3}\pi r^3$ is the areal volume.
This is the unified first law of Ref.\cite{1st}.
One may regard $wdV$ as a type of work and $A\psi$ as an energy supply,
analogous to heat supply $\delta Q=\oint q$,
where $q$ is the heat flux.
The energy supply can be written as
\begin{equation}
A\psi={\kappa dA\over{8\pi}}+rd\left({E\over{r}}\right).
\end{equation}
The second term vanishes when projected along a black-hole horizon,
defined as in Refs.\cite{1st,bhd,ge} by a trapping horizon:
a hypersurface where $dr$ is null,
so that $E=r/2$.
This also occurs for any hypersurface on which $E/r$ is constant,
thereby covering any smooth space-time.
The key point is that the first term is the product of surface gravity $\kappa$
and a total differential.
Identifying $\kappa/2\pi$ as a temperature,
this total differential therefore determines a Clausius entropy $A/4$.
Note that this stems from a purely mathematical property
of the energy supply occuring in the first law.
The restriction to spherical symmetry will be removed 
in forthcoming work\cite{MH}.

Wald\cite{W} also gave a definition of entropy
\begin{equation}
S=2\pi\oint Q|_{\kappa=1}
\end{equation}
where $Q$ henceforth denotes a Noether charge 2-form
obtained from a certain type of Lagrangian.
For Einstein gravity, Iyer \& Wald\cite{IW} found
$Q_{ab}=-\epsilon_{abcd}\nabla^c\xi^d/16\pi$,
where $\epsilon$ is the space-time volume form
and $\xi$ is a generating vector for the diffeomorphisms
which is taken to be the Killing vector of a stationary black hole.
In spherical symmetry,
we propose using the Kodama vector $k$ for $\xi$
to give an alternative definition of the entropy of dynamic
black holes.
This prescription effectively corresponds to also replacing
$\xi$ by $k$
in $S_2$ of Jacobsen et al.\cite{JKM}, cf.\ Koga \& Maeda\cite{KM}.
Then either from the above expression for $Q$
or equivalently from the reduced action\cite{AH} by Wald's method,
\begin{equation}
\oint Q={A\kappa\over{8\pi}}
\end{equation}
where the integral is over a sphere of symmetry.
The same formula holds on a Killing horizon
in terms of the usual stationary quantities,
though it does not seem to have been given explicitly in
previous references\cite{W,IW,JKM,KM}.
Thus the Wald-Kodama entropy is
\begin{equation}
S=A/4.
\end{equation}
So both definitions agree.
The motivations also seem similar,
since Wald's construction involved a first law of black-hole statics
based on perturbations $\delta$ of a stationary solution.
However, this involved effectively taking $\delta\kappa=0$,
leading to the suggestive formula $\delta\oint Q=\kappa\delta A/8\pi$.
This seems questionable since it trivializes the simplest first law
for Schwarzschild black holes, $\delta m=\kappa\delta A/8\pi$,
where $\kappa=1/4m$ and $A=16\pi m^2$.
A way to resolve this is suggested in an accompanying paper\cite{M1}.

This agreement suggests that
some combination of the two methods may be useful in general,
assuming neither stationarity nor Einstein gravity.
In this respect we remark only that
some of the key equations have a similar structure,
with $(\psi,{*}j,E/A,rw,{*}1)$ corresponding to what Iyer \& Wald called
$2(\Theta,J,Q,W\cdot\xi,16\pi X)$, averaged over a sphere.
A full correspondence seems to require a restriction on the type of Lagrangian.

In summary, we have considered two definitions of entropy
for dynamic black holes in spherical symmetry.
For Einstein gravity,
both have the same simple form $A/4$ as for stationary black holes.
This is remarkable given that one might have expected significant complications in dynamic cases.
One of the definitions is a direct analogue
of the original Clausius concept of entropy,
arising from a first law containing an energy-supply term
which equals surface gravity times a total differential.
A statistical or quantum definition of black-hole entropy would also be welcome,
but to our knowledge,
existing methods assume at least instantaneous stationarity, 
e.g.\ Ashtekar et al.\cite{ABCK},
or a lower dimension, e.g.\ Carlip\cite{SC}.

It is also interesting that
both methods formally hold not just on a black-hole horizon,
but anywhere in the space-time, as do all the equations displayed above.
Whether any surface in any space-time should have an entropy
related to its area is arguable,
but this does concur with the entanglement entropy approach\cite{M2}.

\bigskip
Acknowledgements.
SAH is supported by a European Union Science and Technology Fellowship.
SM thanks Professors H. Kodama and W. Israel for their continuing
encouragement. He was supported by JSPS Research Fellowships for Young
Scientists, and this work was supported partially by the Grant-in-Aid
for Scientific Research Fund (No.\ 9809228).
MCA is supported by NSF/JSPS Postdoctoral Fellowship for Foreign
Researchers (No.\ P97198).

\end{document}